\documentclass[twocolumn,showpacs,preprintnumbers,amsmath,amssymb,APSl,prd,nofootinbib,superscriptaddress]{revtex4-1}

\usepackage{dcolumn}
\usepackage{bm}
\usepackage{ifpdf}
\usepackage{hyperref}
\usepackage{bm}
\usepackage{xcolor,color,graphicx,graphics}
\usepackage[spanish,english]{babel}
\usepackage[latin1]{inputenc}
\usepackage[OT1]{fontenc}
\usepackage{latexsym,amssymb,amsmath,amsfonts}
\usepackage{makeidx}
\usepackage{epsfig,subfigure}
\usepackage{natbib}
\usepackage{epstopdf}
\usepackage{mathrsfs}
\usepackage{hyperref}
\hypersetup{colorlinks=true, linkcolor=blue, citecolor=green}
\usepackage{enumerate}

\definecolor{red}{rgb}{1,0,0}

\def\+{^\dagger}

\def\<{\leftarrow}
\def\>{\rightarrow}

\def\({\left(}
\def\){\right)}

\newcommand{\bi}{\begin{itemize}} 				\newcommand{\ei}{\end{itemize}}
\newcommand{\benu}{\begin{enumerate}} 		\newcommand{\enu}{\end{enumerate}}
\newcommand{\bd}{\begin{dinglist}{0}}     \newcommand{\ed}{\end{dinglist}}
\newcommand{\bfig}{\begin{figure}[htbp]}  \newcommand{\efig}{\end{figure}}
        			
\newcommand{\bc}{\begin{center}} 				  \newcommand{\ec}{\end{center}}
\newcommand{\be}{\begin{equation}} 				\newcommand{\ee}{\end{equation}}
\newcommand{\bsub}{\begin{subequations}}  \newcommand{\esub}{\end{subequations}}
\newcommand{\ben}{\begin{eqnarray}} 			\newcommand{\een}{\end{eqnarray}}
\newcommand{\ba}[1]{\begin{array}{#1}} 		\newcommand{\ea}{\end{array}}
\newcommand{\bea}{\begin{equation}\begin{array}{rcl}}
\newcommand{\eea}{\end{array}\end{equation}}

\begin{document}
\title{Parameterized non-relativistic limit of stellar structure equations in Ricci-based gravity theories}

\author{Gonzalo J. Olmo} \email{gonzalo.olmo@uv.es}
\affiliation{Departamento de F\'{i}sica Te\'{o}rica and IFIC, Centro Mixto Universidad de Valencia - CSIC.
Universidad de Valencia, Burjassot-46100, Valencia, Spain}
\affiliation{Departamento de F\'isica, Universidade Federal da
Para\'\i ba, 58051-900 Jo\~ao Pessoa, Para\'\i ba, Brazil}
\author{Diego Rubiera-Garcia} \email{drubiera@ucm.es}
\affiliation{Departamento de F\'isica Te\'orica and IPARCOS, Universidad Complutense de Madrid, E-28040 Madrid, Spain}
\author{Aneta  Wojnar}
\email{aneta.magdalena.wojnar@ut.ee}
\affiliation{Laboratory of Theoretical Physics, Institute of Physics, \\
University of Tartu, W. Ostwaldi 1, 50411 Tartu, Estonia}

\date{\today}

\begin{abstract}
We present the non-relativistic limit of the stellar structure equations of Ricci-based gravities, a family of metric-affine theories whose Lagrangian is built via contractions of the metric with the Ricci tensor of an a priori independent connection. We find that this limit is characterized by four parameters that arise in the expansion of several geometric quantities in powers of the stress-energy tensor of the matter fields. We discuss the relevance of this result for the phenomenology of  non-relativistic stars, such as main-sequence stars as well as several substellar objects.
\end{abstract}

\maketitle

\section{Introduction}

In the last few years our community has witnessed the success of Einstein's General Theory of Relativity (GR) when confronted with a new pool of astrophysical data: from the LIGO and VIRGO observations of gravitational waves out of binary mergers \cite{TheLIGOScientific:2017first,TheLIGOScientific:2017qsa}, to the radio measurements by the EHT of the plasma around the supermassive central object of the M87 galaxy \cite{Akiyama:2019cqa}, both of which are compatible with the predictions of GR on the properties of compact objects (black holes and neutron stars). Together with solar system experiments \cite{Will:2014kxa}, and the compatibility of the concordance cosmological model with observations \cite{Bull:2015stt}, this bunch of predictions have underpinned our trust on the reliability of GR to describe a wide range of gravitational phenomena. However, at the same time, the exploration of the strong-field regime by these new probes also enables us to test alternative descriptions of the gravitational field - commonly known as {\it modified theories of gravity} - via their accompanying phenomenology.  For  some basic references on these theories see e.g.  \cite{Olmo:2011uz,DeFelice:2010aj, CLreview,Clifton:2011jh,Nojiri:2017ncd,BeltranJimenez:2017doy,Heisenberg:2018vsk}.

In addition to black holes and their observational discriminators in terms of gravitational waves \cite{Cardoso:2016oxy,Cardoso:2017cqb,Konoplya:2018yrp,Berti:2018vdi,Ezquiaga:2020dao} and lensing of electromagnetic radiation \cite{Gralla:2019xty,Carballo-Rubio:2018jzw,Wielgus:2020uqz,Afrin:2021imp,Junior:2021atr,Peng:2021osd}, the astrophysics of stellar objects represents another promising playground to test the viability of these modified theories of gravity, thanks to the availability of numerous data from different probes \cite{Berti:2015itd}. From a theoretical point of view, stellar objects are usually split into two classes: relativistic stars such as  neutron stars, where the full power of the Einstein field equations (via the Tolman-Oppenheimer-Volkoff equation) must be called upon \cite{Lattimer:2004pg}; and non-relativistic stars, to which main-sequence stars as well as brown and red dwarfs and giant exoplanets \cite{Burrows:1992fg} belong\footnote{While white dwarfs are typically classified as non-relativistic objects, relativistic effects need to be incorporated to reliably describe certain aspects of the physics of these stars \cite{GlenBook}.}, where one can neglect the contribution of the pressure of the matter fields as compared to their energy density.

As opposed to neutron stars, where the uncertainty of the equation of state at supranuclear densities is entangled with the new parameter(s) of the modified gravity side (for a review of this issue see \cite{Olmo:2019flu}), non-relativistic stars may offer a cleaner scenario to test modifications to GR predictions since they are more weakly dependent on non-gravitational physics than their relativistic counterparts. This is of particular relevance for those models in which the new physics appears as a modification of the Poisson equation inside astrophysical bodies \cite{Saito:2015fza}. Despite the fact that numerous additional aspects of the modeling of such stars (including thermodynamics, description of their atmospheres, energy generation rates, etc.) are needed in order to achieve a realistic description of their structure, simple analytical models are capable to provide reasonable estimates to some global properties of such non-relativistic stars. Examples of this are the limiting masses, such as the Chandrasekhar mass for white dwarfs \cite{Chandra,Saltas:2018mxc,Jain:2015edg,Banerjee:2017uwz,Wojnar:2020wd,Belfaqih:2021jvu} or the minimum required mass for a star to stably burn hydrogen \cite{Sakstein:2015zoa,Sakstein:2015aac,Crisostomi:2019yfo} and deuterium \cite{Rosyadi:2019hdb}, but it may also have a non-negligible impact in the description of the early evolution of low-mass stars \cite{Wojnar:2020txr}, in the cooling process of brown dwarfs \cite{Benito2021}, and in age-estimation techniques such as those based on the lithium depletion method \cite{Wojnar:2020frr}.

The main aim of the present paper is to work out the non-relativistic limit of the stellar structure equations corresponding to a large family of metric-affine theories of gravity, where metric and affine connection are regarded as independent entities \cite{Bahamonde:2021akc}. More specifically, our target are those theories built as scalar functions of the metric and the Ricci tensor of the affine connection (Ricci-based gravities, or RBGs for short \cite{Afonso:2018bpv}). The interest in this family of theories lies on the fact that their vacuum dynamics is exactly that of GR (plus a cosmological constant term, which is usually neglected in stellar structure models), which allows (most of) them to automatically pass both solar system experiments and gravitational wave observations so far, while the non-trivial role played by the connection induces new non-linearities engendered by the matter fields inside astrophysical bodies \cite{Rubiera-Garcia:2020gcl}. This has important consequences for the predictions of these theories regarding the above aspects of non-relativistic stars, as illustrated by some applications of specific theories recently worked out in the literature, such as the minimum main sequence mass of quadratic $f(R)$ gravity \cite{Olmo:2019qsj}, or the minimum deuterium burning mass for Eddington-inspired Born-Infeld gravity (EiBI, \cite{Rosyadi:2019hdb}), which have allowed to place constraints on the parameters of these theories.

In this work we shall show that the non-relativistic limit of the whole RBG family can be fully characterized in terms of four parameters associated to the expansion in powers of the stress-energy tensor of the matter fields, two of which belong to the shape of the RBG Lagrangian density, and the other two to the deformation matrix that relates the  Einstein and RBG frame metrics of the corresponding field equations. From the general expression that we obtain, those of $f(R)$ and EiBI gravity are easily obtained, which provides a consistency check of our results. We shall  discuss some prospects for applications of this non-relativistic limit of RBGs inside astrophysical bodies.

\section{Ricci-based gravity theories}

\subsection{Action and field equations}

The class of RBG theories is defined by the action
\begin{equation} \label{eq:actionRBG}
\mathcal{S}=\int d^4 x \sqrt{-g} \mathcal{L}_G(g_{\mu\nu},R_{\mu\nu})
+ \mathcal{S}_m(g_{\mu\nu},\psi_m)  \ ,
\end{equation}
where $g$ is the determinant of the space-time metric $g_{\mu \nu}$, the Ricci tensor $R_{\mu \nu}$ is built out solely with the affine connection $\Gamma \equiv \Gamma_{\mu\nu}^{\lambda}$, which is independent of the metric (Palatini or metric-affine approach), while the gravitational Lagrangian $\mathcal{L}_G$ is a scalar function built out of powers of traces of the object ${M^ \mu}_{\nu} \equiv g^{\mu \alpha}R_{\alpha\nu}$. As for the matter sector, $\mathcal{S}_m=\int d^4 x \sqrt{-g} \mathcal{L}_m(g_{\mu\nu},\psi_m)$, it is assumed to be minimally coupled to the metric, with $ \psi_m$ denoting collectively the matter fields. In this action we have neglected the role of torsion (the antisymmetric part of the connection), since for minimally coupled bosonic fields it trivializes \cite{Afonso:2017bxr}, and the antisymmetric part of the Ricci tensor, to safeguard the theory against potential ghost-like instabilities \cite{BeltranJimenez:2019acz,Jimenez:2020dpn}. Defined this way, the RBG family  includes some well known theories, such as GR itself, $f(R)$ gravity, quadratic gravity, as well as EiBI gravity and many of its extensions \cite{BeltranJimenez:2017doy}.

It has been shown elsewhere (see for instance Ref.\cite{BeltranJimenez:2017doy}) that the field equations of the action (\ref{eq:actionRBG}) admit a representation of the form
\begin{equation} \label{eq:feRBG}
{G^\mu}_{\nu}(q)=\frac{\kappa^2}{\vert \hat{\Omega} \vert^{1/2}} \left({T^\mu}_{\nu}-\delta^\mu_\nu \left(\mathcal{L}_G + \frac{T}{2} \right) \right) \ ,
\end{equation}
where $T_{\mu\nu}=-\frac{2}{\sqrt{-g}}\frac{\partial \mathcal{L}_m}{\partial g^{\mu\nu}}$ is the energy-momentum tensor of the matter fields with $T$ its trace, while ${G^\mu}_{\nu}(q)$ is the Einstein tensor associated to a new rank-two tensor $q_{\mu\nu}$ such that the connection $\Gamma$ is Levi-Civita of it, that is, $\nabla_{\mu}^{\Gamma}(\sqrt{-q} q^{\alpha \beta})=0$. For any RBG this tensor is related to the space-time metric $g_{\mu\nu}$ via the fundamental relation
\begin{equation}\label{eq:defmat}
q_{\mu\nu}=g_{\mu\alpha}{\Omega^\alpha}_{\nu} \ ,
\end{equation}
where the \emph{deformation matrix} ${\Omega^\alpha}_{\nu}$ (whose determinant is denoted by vertical bars) depends on the particular $\mathcal{L}_G$ chosen, but it can always be expressed on-shell (as so does $\mathcal{L}_G$ itself) as a function of the matter fields and (possibly) the space-time metric too. The representation (\ref{eq:feRBG}) also reinforces the fact that the RBG field equations are second-order and that in vacuum, ${T_\mu}^{\nu}=0$, these equations and their solutions boil down to their GR counterparts. Therefore, no extra degrees of freedom (beyond the usual two polarizations of the gravitational field) are propagated in these theories.

\subsection{The two frames of RBGs and the mapping equations}

The resemblance of the representation (\ref{eq:feRBG}) with the usual GR field equations strongly suggests that an Einstein frame should also be possible, that is \cite{JB}
\begin{equation} \label{eq:feGR}
{G^\mu}_{\nu}(q)=\kappa^2 \tilde{T}{^\mu}_{\nu}(q) \ ,
\end{equation}
where $\tilde{T}{^\mu}_{\nu}(q)$ would be the energy-momentum tensor of some new matter Lagrangian density $\tilde{\mathcal{L}}_m (q_{\mu\nu},\tilde{\psi}_m)$, built of the same type of fields as the original ${T^\mu}_{\nu}(g)$, but with a different functional dependence. For this representation to work, one needs to systematically remove all dependencies on the space-time metric $g_{\mu\nu}$ in favour of those of $q_{\mu\nu}$. This intuitive idea was given explicit support and internal consistence in a series of works by some of us \cite{Afonso:2018bpv,Afonso:2018mxn,Afonso:2018hyj}. The most general case for the matter sector studied so far considers an anisotropic fluid (which is suitable for stellar structure modelling), whose energy-momentum tensor can be written, in terms of the space-time metric, as
\begin{equation} \label{eq:Tmunuani}
{T^\mu}_{\nu}(g)=\left(\rho+p_{\perp}\right) u^{\mu} u_{\nu}+p_{\perp} \delta_{\nu}^{\mu}+\left(p_{r}-p_{\perp}\right) \chi^{\mu} \chi_{\nu} \ ,
\end{equation}
where $\rho$ is the energy density, and $p_r$ and $p_{\perp}$  the radial and tangential pressures, respectively, and we have introduced the unit  time-like  $g^{\mu\nu}u_\mu u_\nu=-1$ and space-like $g^{\mu\nu}\chi_\mu \chi_\nu=+1$ vectors, respectively. Note that in a comoving frame this energy-momentum tensor can simply be expressed as ${T^\mu}_\nu=\text{diag}(-\rho,p_r,p_{\perp},p_{\perp})$.

In the GR frame, the corresponding energy-momentum tensor reads formally the same
\begin{equation}
\tilde{T}{^\mu}_{\nu}(q)=\left(\tilde\rho+\tilde p_{\perp}\right) v^{\mu} v_{v}+\tilde p_{\perp} \delta_{\nu}^{\mu}+
\left(\tilde p_{r}-\tilde p_{\perp}\right) \xi^{\mu} \xi_{v} \ ,
\end{equation}
with new energy density $\tilde{\rho}$ and radial $\tilde{p}_r$ and tangential $\tilde{p}_\perp$ pressures, and unit time-like  $q^{\mu\nu}v_\mu v_\nu=-1$ and space-like $q^{\mu\nu}\xi_\mu \xi_\nu=+1$ vectors. Comparing the representations (\ref{eq:feRBG}) and  (\ref{eq:feGR}) of the field equations one finds that the energy density and pressures of the two fluids must be related as
\begin{eqnarray}
\tilde{p}_{\perp}&=&\frac{1}{\vert \hat{\Omega} \vert^{1/2}} \left[\frac{\rho-p_r}{2} -\mathcal{L}_G \right] \label{eq:mapflu1} \\
\tilde{\rho}+\tilde{p}_{\perp}&=&\frac{\rho+p_{\perp}}{\vert \hat{\Omega} \vert^{1/2}}  \label{eq:mapflu2} \\
\tilde{p}_r-\tilde{p}_{\perp}&=&\frac{p_r-p_{\perp}}{\vert \hat{\Omega}  \vert^{1/2}}  \ , \label{eq:mapflu3}
\end{eqnarray}
which shall be called hereafter the \emph{mapping equations} between the two frames. Given that the determinant of the deformation matrix, $\vert \hat{\Omega}  \vert \equiv \vert \hat{\Omega}  \vert  ({T^\mu}_\nu,g_{\mu\nu}) $, is a function of $\rho$, $p_r$, and $p_\perp$, these equations establish a correspondence between GR coupled to some anisotropic fluid (tilded variables) and a given RBG coupled to another anisotropic fluid (untilded variables). Moreover, this correspondence can be used to obtain solutions of the latter starting from a seed solution of the former using purely algebraic transformations, as explicitly verified in Refs. \cite{Afonso:2018mxn,Afonso:2018hyj} with concrete examples. Besides their obvious interest for the sake of finding analytic solutions of physical interest of modified gravity \cite{Guerrero:2020azx}, the mapping equations also allow us to carry out formal manipulations to handle specific systems, as the one of (the newtonian limit of) stellar structure considered  in this work.

\section{Parameterized non-relativistic limit of RBGs}

For the analysis of non-relativistic stars (and relativistic stars alike) one typically neglects anisotropies, whose estimated effects in realistic scenarios are (assumed to be) smaller than other degeneracies/unknowns (see however \cite{Raposo:2018rjn}). Thus we can set $p_{\perp}=p_r=p$ and $\tilde{p}_{\perp}=\tilde{p}_r=\tilde{p}$  to obtain a perfect fluid on both frames:
\begin{eqnarray} \label{eq:pf}
{T^\mu}_{\nu}&=&\left(\rho+p\right) u^{\mu} u_{\nu}+p\delta_{\nu}^{\mu} \\
 \tilde{T}{^\mu}_{\nu}&=&\left(\tilde\rho+\tilde p\right) v^{\mu} v_{v}+\tilde p \delta_{\nu}^{\mu} \ ,
\end{eqnarray}
such that the mapping equations (\ref{eq:mapflu1}), (\ref{eq:mapflu2}) and (\ref{eq:mapflu3}) boil down to
\begin{eqnarray}
\tilde{p}&=&\frac{1}{\vert {\Omega} \vert^{1/2}} \left[ \frac{\rho-p}{2} - \mathcal{L}_G \right] \label{eq:map1b} \\
\tilde{\rho} + \tilde{p} &=& \frac{\rho + p }{\vert {\Omega} \vert^{1/2}} \ . \label{eq:map2b}
\end{eqnarray}
In order to proceed further, we must find a representation of the deformation matrix ${\Omega^\mu}_{\nu}$. Since this quantity is a nonlinear function of the energy-momentum tensor, it can be formally written as 
\begin{equation}
{\Omega^\mu}_{\nu}=\delta^\mu_\nu + \sum_{n=1}^\infty \sum_{i_n=0}^3 c_{(n,i_n)}{(T^{(n,i_n)})^\mu}_\nu
\end{equation}
where the ${(T^{(n,i_n)})^\mu}_\nu$ represent tensor structures that can be built with $n$ powers of the energy-momentum tensor \cite{BeltranJimenez:2021iqs}. The zeroth order term $\delta^\mu_\nu$ simply indicates that in vacuum one recovers the GR dynamics. Given the simplicity of (\ref{eq:map1b}) and (\ref{eq:map2b}), it is easy to see that these infinite sums must boil down to  
\begin{equation}
{\Omega^\mu}_{\nu}=A\delta^\mu_\nu + Bu^\mu u_\nu \ ,
\end{equation}
using the RBG frame variables, with $A$ and $B$ being some functions of $\rho$ and $p$, or to
\begin{equation}
{\Omega^\mu}_{\nu}=\tilde{A}\delta^\mu_\nu + \tilde{B}v^\mu v_\nu
\end{equation}
if we employ the GR frame variables instead, with $\tilde{A}$ and $\tilde{B}$ some functions of $\tilde{\rho}$ and $\tilde{p}$. Focusing on the zeroth and first order contributions, which will be the dominant terms in a perturbative expansion, we get  
\begin{eqnarray} \label{eq:Omeexpan}
{\Omega^\mu}_{\nu} &\approx& \delta^\mu_\nu + \alpha T \delta^\mu_\nu + \beta{T^\mu}_{\nu} \nonumber \\
&=&(1+\alpha(3 p-\rho)+\beta p)\delta^\mu_\nu + \beta(\rho+p) u^\mu u_\nu \ ,
\end{eqnarray}
where we have made use of the equality $T=-\rho+3p$ and suitably rearranged terms.  In the non-relativistic limit, $p \ll \rho$, the components of this matrix read explicitly as ($i=1,2,3$)
\begin{equation} \label{eq:Omcom}
{\Omega^t}_t=1-(\alpha +\beta) \rho \hspace{0.2cm} ; \hspace{0.2cm} {\Omega^i}_i=1-\alpha \rho \ ,
\end{equation}
and hereafter we shall refer to this limit and neglect contributions in $p$ (unless explicitly stated) as compared to those in $ \rho$ (at the same order). Since this matrix is diagonal, the computation of the determinant $\vert \hat{\Omega} \vert$ is straightforward:
\begin{equation} \label{eq:detn}
\vert \Omega\vert^{1/2}= ({\Omega^t}_t)^{1/2}({\Omega^i}_i)^{3/2} = 1-\left(2\alpha + \frac{\beta}{2}\right) \rho \ .
\end{equation}
Let us point out that similar expressions for ${ \Omega^\mu}_{\nu}$ and its determinant can be written in the GR frame (tilted variables) just by adding tildes to all the constants and functions. Then the mapping equations (\ref{eq:map1b}) and (\ref{eq:map2b}) will provide the correspondence between the functions in both representations. Using one of them or the other is thus just a matter of convenience for each specific computation, though at the end of the day the physical result has to be written in terms of the physical variables (that is, the untilded ones).

Now we need to find a similar expansion for the gravitational Lagrangian $\mathcal{L}_G$. Since it is a scalar quantity, in general, it can be expanded as an infinite series of 
traces of the objects ${(T^{(n,i_n)})^\mu}_\nu$. Likewise for ${\Omega^\mu}_{\nu}$, retaining terms up to quadratic order in this series yields the result\footnote{We are neglecting here the zeroth-order term, which would correspond to a cosmological constant that does not contribute significantly to stellar quantities.}
\begin{equation} \label{eq:LGex}
\mathcal{L}_G=-\frac{1}{2\kappa^2}\left( \kappa^2 T + \gamma {T^\mu}_{\alpha}{T^\alpha}_{\mu} + \delta T^2 \right) \ ,
\end{equation}
where we have inserted a factor $\kappa^2$ in the first term to identify it as the GR relation between the curvature scalar and the trace of the energy-momentum tensor. The other two terms yield the RBG corrections at quadratic order in the energy-momentum tensor and its trace, where we stress that for dimensional consistency the parameters $(\gamma,\delta)$ must have implicit $\kappa^4$ factors. For the perfect fluid energy-momentum tensor (\ref{eq:pf}) this expression becomes, for non-relativistic fields:
\begin{equation}
\mathcal{L}_G \approx \frac{\rho-3p}{2} - \frac{(\gamma+\delta)\rho^2}{2\kappa^2} \ ,
\end{equation}
where for the moment we have kept the term linear in $p$ but neglected one in $\rho \cdot p$  and another in $p^2$, as they are assumed to be smaller than the one quadratic in $\rho$.

Having under control the main objects on the right-hand-side of the mapping equations (\ref{eq:map1b}) and (\ref{eq:map2b}), the next step in our analysis is to cast the field equations for stellar structure (the TOV equations). From the GR representation (\ref{eq:feGR}),  such equations are found from the conservation of the energy-momentum tensor (as follows from the fulfillment of Bianchi's identities in (\ref{eq:feGR})), $\nabla_{\mu} \tilde{T}{^\mu}_{\nu}(q)=0$, which, for a static, spherically symmetric metric, $ds_q^2=q_{tt}dt^2 + q_{xx}dx^2+x^2d\Omega^2$, yields
\begin{eqnarray}
\tilde{p}_{x}=-(\tilde{\rho}+\tilde{p}) \frac{q^{tt}}{2} \partial_{x} q_{tt} \ .  \label{eq:Prq}
\end{eqnarray}
Furthermore, by assuming this line element to take the form
\begin{equation}
ds_q^2=-C(x)e^{\psi(x)}dt^2+\frac{dx^2}{C(x)} + x^2 d\Omega^2 \ ,
\end{equation}
and introducing the standard mass ansatz $C(x)=1-2M(x)/x$, then the functions $\{\psi(x)$, $M(x)\}$ can be determined by resolution of the RBG field equations (\ref{eq:feGR}). Such a resolution parallels the analog problem in GR, and yields the well known result
\begin{equation} \label{eq:NRLq}
\tilde{p}_x=-\frac{(\tilde{\rho}+\tilde{p})x}{2C(x)} \left[\tilde{p}+\frac{2M(x)}{x^3}\right] \ ,
\end{equation}
which is nothing but the TOV equations in the GR frame, where the mass function satisfies
\begin{equation} \label{eq:massx}
M_x=\frac{\kappa^2 x^2}{2} \tilde{\rho}(x) \ .
\end{equation}
Note that the non-relativistic limit of Eq.(\ref{eq:NRLq}) does not allow us to directly take $\tilde{p}$ away, since it contains terms in $\rho^2$ when moving to the GR frame via the mapping equations (\ref{eq:map1b}) and (\ref{eq:map2b}), which is our next goal. To this end, upon suitable combinations of these mapping equations, together with Eq.(\ref{eq:detn}), we find in this case the following relations between the energy densities and the pressures on both frames
\begin{eqnarray}
\tilde{\rho}\approx \frac{\rho-\frac{(\gamma+\delta)\rho^2}{2\kappa^2}}{\vert \hat{\Omega} \vert^{1/2}} &\approx& \rho+\left(2\alpha+\frac{\beta}{2}-\frac{\gamma+\delta}{2\kappa^2}\right) \rho^2, \label{eq:rhotilde} \\
\tilde{p}\approx \frac{p+\frac{(\gamma+\delta)\rho^2}{2\kappa^2}}{\vert \hat{\Omega} \vert^{1/2}} &\approx & p+\frac{(\gamma+\delta)}{2\kappa^2}\rho^2 \ . \label{eq:ptilde}
\end{eqnarray}
 On the other hand, we need to incorporate in this analysis the explicit transformation between the radial coordinates in the GR and RBG frames via the fundamental relation (\ref{eq:defmat}). Thus, assuming also a static, spherically symmetric line element for the space-time geometry $g_{\mu\nu}$ with a new radial coordinate $r$, using (\ref{eq:Omcom}) one finds that, in the non-relativistic approximations employed so far, Eq.(\ref{eq:defmat}) yields the result
\begin{equation} \label{eq:coordi}
x^2=r^2\left(1-\alpha \rho\right) \ ,
\end{equation}
neglecting again both $\rho^2$ and $\rho \cdot P$ terms in this approximation.

We are now ready to  express  (\ref{eq:NRLq}) in terms of the physical variables. We start with the mass function (\ref{eq:massx}). Using (\ref{eq:rhotilde}) and (\ref{eq:coordi}) one finds that the mass function can be integrated as
\begin{eqnarray} \label{eq:Mint}
M(r) \approx M_0(r) + \eta(r) \ ,
\end{eqnarray}
where we have introduced the definition
\begin{equation}
M_0(r)=\frac{\kappa^2}{2} \int^r dR R^2 \rho(R) \ ,
\end{equation}
which is nothing but the GR contribution to the mass, while the additional piece reads (here $\rho' \equiv d\rho/dr$)
\begin{equation} \label{eq:eta}
\eta(r)=\frac{\kappa^2}{4} \int^r dR R^2 \rho^2 \left[ \alpha \left(1-\frac{r\rho'}{\rho}\right) + \beta  -\frac{\gamma+\delta}{\kappa^2} \right] \ ,
\end{equation}
in which we have neglected again cubic terms in $\rho$ in the integrand. Now, from  Eq.(\ref{eq:NRLq}), using the expression for the mass function (\ref{eq:Mint}) alongside the mapping equations particularized to this case, Eqs.(\ref{eq:rhotilde}) and (\ref{eq:ptilde}), we get
\begin{eqnarray}
\tilde{p}_x &\approx& -\frac{r\rho}{2}\Big[ \Big(p+\frac{2M_0}{r^3} \Big) + \frac{2M_0}{r^3} \Big( \frac{6\alpha+\beta}{2} \Big) \rho + \nonumber \\
&+&\Big( \frac{\gamma+\delta}{2\kappa^2} \Big)\rho^2 + \Big(1+\frac{3}{2} \alpha \rho \Big)\eta(r)
\Big].
\end{eqnarray}
Finally we need to work out the expression on the left-hand side of this equations in terms of the physical variables using again the relation between coordinates in both frames, Eq.(\ref{eq:coordi}). After suitably rearranging terms, we arrive to the final result
\begin{eqnarray} \label{master equation}
p_r &\approx& -\frac{M_0\rho}{r^2}\Big[1+\frac{pr^3}{2M_0}+\Big( \frac{5\alpha + \beta -\alpha r^2 \rho'/\rho}{2}\Big)\rho \\
&+& \frac{(\gamma+\delta) r^3}{4\kappa^2 M_0} \left(1+4\rho'/(r\rho^2) \right)\rho^2  + \frac{r^3}{2M_0} \Big(1+\frac{3}{2} \alpha \rho \Big)\eta(r) \Big]. \nonumber
\end{eqnarray}
This equation fully parameterizes the non-relativistic limit of the RBG family in terms of the coefficients $(\gamma,\delta)$, associated to the  expansion of the RBG Lagrangian, $\mathcal{L}_G$, and the coefficients $(\alpha,\beta)$, associated to the expansion of the deformation matrix, ${\Omega^\mu}_{\nu}$. In GR, $\alpha=\beta=\gamma=\delta=0$ and neglecting the pressure term in (\ref{master equation}) one recovers the  Newtonian limit
\begin{equation}
p_r^{GR}=-\frac{M_0(r) \rho(r)}{r^2} \ ,
\end{equation}
from which the well known Lane-Emden equation is derived assuming a polytropic fluid and on top of which studies of non-relativistic stars are typically carried out.

The bottom line of this result is that all RBGs, no matter the explicit functional dependencies of their Lagrangian densities, have the same qualitative behaviour in what concerns their non-relativistic limit of stellar structure. This means that, once a particular RBG model is given, a simple expansion of its Lagrangian density in (\ref{eq:LGex}) and of the deformation matrix in (\ref{eq:Omeexpan}) allows us to find the parameters associated to the non-relativistic regime and, therefore, starting from it different aspects of the physics of non-relativistic stars can be immediately implemented. To this end, one needs to estimate the relative strength of each term contributing to this equation for any kind of such star. Therefore, we first keep in this expression only those terms linear and quadratic in the energy density and linear in the pressure, so we get
\begin{eqnarray} \label{eq:p_rsimplified}
p_r &\approx& -\frac{M_0\rho}{r^2}\Big[1+\frac{pr^3}{2M_0}+\Big( \frac{5\alpha + \beta -\alpha r^2 \rho'/\rho}{2}\Big)\rho \nonumber \\
& +&   \frac{(\gamma+\delta) r^2\rho'}{\kappa^2 M_0})  \Big] \ .
\end{eqnarray}
For a non-relativistic star typically the pressure is modeled by using a polytropic equation of state \cite{GlenBook}
\begin{equation}
p=K  \rho^{\Gamma} \ ,
\end{equation}
where the polytropic constant $K$ and index $\Gamma=1+1/n$ with $n>1$ are capable to characterize different types of non-relativistic stars\footnote{For equations of state $p=(\Upsilon-1)\rho$ with $0<\Upsilon<1$ the pressure term is of order below than the one in $\rho$. However, such equations of interest are not of interest for the physics of these stars and can therefore be neglected in our analysis.}. Therefore, the term in $p$ can be typically neglected as compared to those in $\rho$, and in the non-relativistic limit  the approximations $p \ll \rho, r^3p \ll M, 2M_0/r \ll 1$ hold. Therefore, we shall use formula (\ref{eq:p_rsimplified}) with the constraint above for the discussion on specific examples of RBGs.

\subsection{Quadratic $f(R)$ gravity}

The simplest example of an RBG is given by $f(R)$ theories. Let us consider the well known quadratic (Starobinsky) $f(R)$ gravity, which is given by the Lagrangian density \cite{Staro}
\begin{equation}
f(R)=\frac{1}{2\kappa^2} \left(R+\mu R^2\right) \ ,
\end{equation}
where $\mu$ is a parameter with dimensions of length squared. For this theory, in the metric-affine formalism it turns out that $R=-\kappa^2 T$ via the corresponding field equations (whose trace implies that $Rf_R-2f=\kappa^2 T$), which is the same result as in GR. This allows us to write this Lagrangian in terms of energy density and pressure (in the non-relativistic limit) as
\begin{equation}
\mathcal{L}_G =\frac{1}{2\kappa^2}f(R) \approx  \frac{\rho-3p}{2}+\frac{\mu\kappa^2\rho^2}{2}
\end{equation}
On the other hand, the deformation matrix in this case reads ${\Omega^\mu}_{\nu}=f_R \delta^\mu_{\nu}$, where $f_R \equiv df/dR$, which entails a conformal relation between the space-time and auxiliary metrics. From Eqs.(\ref{eq:rhotilde}) and (\ref{eq:ptilde}), one is equipped with the following relations between energy density and pressure in the GR and RBG frames
\begin{eqnarray}
\tilde{\rho}&=&\frac{\rho-\mu \kappa^2/4\rho^2}{(1+\epsilon \kappa^2 \rho)^2}\approx \rho - \frac{7\mu\kappa^2\rho^2}{2} \ ,\\
\tilde{p}&=&\frac{p+\mu \kappa^2/4\rho^2}{(1+\epsilon \kappa^2 \rho)^2}\approx p +
\frac{\mu\kappa^2\rho^2}{2} \ ,
\end{eqnarray}
which agrees with the results of  \cite{Mana:2015vqa}. With these expressions at hand, and from Eqs.(\ref{eq:Omcom}) and (\ref{eq:LGex}), it is a simple exercise to see that the parameters $\alpha=-2\mu \kappa^2,\beta=0,\gamma=-\mu \kappa^4, \delta=0$ and thus Eq.(\ref{eq:p_rsimplified}), neglecting the contribution of the pressure, reads
\begin{equation}
 p_r =
-\frac{M_0\rho}{r^2}\left[1-\mu\kappa^2 (5\rho-r\rho') - \frac{\mu\kappa^2 r^2 \rho'}{M_0}  \right]
\end{equation}
which is the expression for the non-relativistic limit of quadratic $f(R)$ gravity with the assumptions above.

\subsection{Eddington-inspired Born-Infeld gravity}

Another well known example of a RBG employed in the literature is EiBI gravity, given by the action \cite{banados}
\begin{equation}
S_{EiBI}=\frac{1}{ \epsilon \kappa^2} \int d^4x \left[\sqrt{\vert g_{\mu\nu}+\epsilon R_{\mu\nu} \vert}-\lambda \sqrt{-g} \right]
\end{equation}
where $\epsilon$ is EiBI parameter with dimensions of length squared, and the theory enjoys an effective cosmological constant $\Lambda_{eff}=\frac{\lambda-1}{\epsilon}$. The astrophysical and cosmological phenomenology for this theory and its many extensions was thoroughly discussed in \cite{BeltranJimenez:2017doy}. The Lagrangian density for EiBI gravity can be conveniently written as
\begin{equation} \label{eq:LBI}
\mathcal{L}_{EiBI}=\frac{\vert \hat{\Omega}\vert^{1/2}  - \lambda}{\epsilon \kappa^2} \ ,
\end{equation}
while the deformation matrix in this case is implicitly given by the algebraic equation
\begin{equation}
\vert \hat{\Omega}\vert^{1/2} ({\Omega^\mu}_{\nu})^{-1}=\lambda \delta^\mu_\nu - \epsilon \kappa^2 {T^\mu}_{\nu} \ .
\end{equation}
In order to solve it, one inserts the expression of the perfect fluid (\ref{eq:pf}) and upon resolution and comparison with (\ref{eq:Omcom}) one finds (for asymptotically flat configurations, $\lambda=1$) the parameters $(\gamma+ \delta)=\epsilon \kappa^2/4$ while from Eq.(\ref{eq:LBI}) one gets that  $\alpha=-\epsilon \kappa^2/2, \beta=\epsilon \kappa^2$. Therefore, upon substitution in Eq.(\ref{eq:p_rsimplified}), we are left with the result
\begin{equation}
p_r=-\frac{M_0\rho}{r^2}   - \frac{\epsilon \kappa^2 \rho \rho'}{4}  +\frac{\epsilon \kappa^2 M_0  (3-r\rho'/\rho)\rho}{4r^2} \ ,
\end{equation}
and then by imposing that $r \gg 2M_0$ the last term can be neglected and one ends up with the expression reported in Ref.\cite{vitor}.

\section{Conclusion and discussion}

The physics of non-relativistic stars, though comparatively much less explored than their relativistic  counterparts (neutron stars), nonetheless offers a viable opportunity to test the predictions and consistency of modified gravity theories with astrophysical observations. It is so because of less severe unknowns on non-gravitational aspects of these stars, since those can be well modeled from a  phenomenological perspective via observations. Though fully reliable predictions can only be obtained after incorporating all the physics known to play a relevant role in these objects and require the use of numerical simulations, simple analytical models offer reasonable approximations to some of the most relevant global aspects of the physics of non-relativistic stars.

The above scenario is particularly relevant for those modified theories of gravity yielding  modifications of the gravitational field inside astrophysical sources while leaving the vacuum dynamics unaffected (so as to pass solar system experiments). This is a property naturally hold by the family of Ricci-based gravities considered in this work, without any need of implementing screening/chamaleon mechanisms \cite{Crisostomi:2019yfo}. Elaborating from the field equations of these theories framed in a representation in terms of an auxiliary metric (its Einstein frame), we have used a recently found mapping between these equations and those of GR, by which both theories are coupled to the same kind of fields but with different functional dependencies on their Lagrangian densities, the explicit form of which is obtained once a specific RBG theory is given. Here we have shown that this mapping allows formal manipulations of the corresponding field equations simplifying the process to obtain the corresponding non-relativistic limit of the stellar structure equations, somewhat paralleling the well known duality of the Einstein/Jordan frames of the metric $f(R)$ case. Via this process, the main result of the present work is to show that all RBGs enjoy the same non-relativistic equation, parameterized in terms of four constants associated to the expansion in powers of the energy-momentum of the matter fields of the gravitational Lagrangian and of the deformation matrix realizing the transformation between the space-time and auxiliary metrics.

Many recent works in the field have shown the viability of the non-relativistic stellar limit of modified theories of gravity to obtain specific predictions of these theories that can be accessible via astronomical observations. Prominent among them are the limiting masses between different kinds of dwarf and main-sequence stars  \cite{Saltas:2018mxc,Jain:2015edg,Banerjee:2017uwz,Belfaqih:2021jvu,Wojnar:2020wd,Sakstein:2015zoa,Sakstein:2015aac,Crisostomi:2019yfo,Rosyadi:2019hdb,Wojnar:2020txr,Benito2021,Wojnar:2020frr,Olmo:2019qsj}, including some recent challenges to these limiting masses such as the existence of super-Chandrasekhar white dwarfs, whose masses are suggested to be up to two times the standard Chandrasekhar limit \cite{Howell:2006vn,HaKaSaNo,Das:2013gd,Hsiao:2020whc}, a finding that could be explained via modified gravity \cite{Wei:2021xek}. The RBG family of theories and its non-relativistic limit introduced here therefore offer a suitable playground to keep exploring the predictions of theories when supplemented with suitable equations of state and incorporating the additional physical elements needed to reliably model dwarf and main-sequence stars. This would allow to place more stringent constraints on the astrophysical viability of these theories as compared to GR. 

\section*{Acknowledgments}
 DRG is funded by the \emph{Atracci\'on de Talento Investigador} programme of the Comunidad de Madrid (Spain) No. 2018-T1/TIC-10431, and acknowledges further support from the Ministerio de Ciencia, Innovaci\'on y Universidades (Spain) project No. PID2019-108485GB-I00/AEI/10.13039/501100011033, and the FCT projects No. PTDC/FIS-PAR/31938/2017 and PTDC/FIS-OUT/29048/2017. A. W. is supported by the EU through the European Regional Development Fund CoE
program TK133 ``The Dark Side of the Universe". This work is supported by the Spanish projects  FIS2017-84440-C2-1-P (MINECO/FEDER, EU), PROMETEO/2020/079 (Generalitat Valenciana), and i-COOPB20462 (CSIC), and by the Edital 006/2018 PRONEX (FAPESQ-PB/CNPQ, Brazil, Grant 0015/2019). This article is based upon work from COST Action CA18108, supported by COST (European Cooperation in Science and Technology).
DRG and AW thank the Department of Physics and IFIC of the University of Valencia for their hospitality during different stages of the elaboration of this work.

\end{document}